\documentclass[showpacs,prl,superscriptaddress,preprint]{revtex4}
\usepackage{amssymb}
\usepackage{graphicx}
\usepackage{endnotes}

\begin{document}

\title{Electron-Phonon Interactions in Bilayer Graphene: A First Principles Approach}

\author{K. M. Borysenko}
\affiliation{Department of Electrical and Computer Engineering,
North Carolina State University, Raleigh, NC 27695-7911}

\author{J. T. Mullen}
\affiliation{Department of Physics, North Carolina State University,
Raleigh, NC 27695-8202}

\author{X. Li}
\affiliation{Department of Electrical and Computer Engineering,
North Carolina State University, Raleigh, NC 27695-7911}

\author{Y. G. Semenov}
\affiliation{Department of Electrical and Computer Engineering,
North Carolina State University, Raleigh, NC 27695-7911}

\author{J. M. Zavada}
\affiliation{Department of Electrical and Computer Engineering,
North Carolina State University, Raleigh, NC 27695-7911}

\author{M. Buongiorno Nardelli}
\affiliation{Department of Physics, North Carolina State University,
Raleigh, NC 27695-8202} \affiliation{CSMD, Oak Ridge National
Laboratory, Oak Ridge, TN 37831}

\author{K. W. Kim}\email{kwk@ncsu.edu}
\affiliation{Department of Electrical and Computer Engineering,
North Carolina State University, Raleigh, NC 27695-7911}

\begin{abstract}

Density functional perturbation theory is used to analyze
electron-phonon interaction in bilayer graphene. The results show
that phonon scattering in bilayer graphene bears more resemblance
with bulk graphite than monolayer graphene. In particular,
electron-phonon scattering in the lowest conduction band is
dominated by six lowest (acoustic and acoustic-like) phonon branches
with only minor contributions from optical modes. The total
scattering rate at low/moderate electron energies can be described
by a simple two-phonon model in the deformation potential
approximation with effective constants $D_{ac} \approx 15$~eV and
$D_{op} \approx 2.8\times10^8$~eV/cm for acoustic and optical
phonons, respectively. With much enhanced acoustic phonon
scattering, the low field mobility of bilayer graphene is expected
to be significantly smaller than that of monolayer graphene.

\end{abstract}








\maketitle


Graphene, a two-dimensional sheet of carbon atoms in a honeycomb
lattice, has received wide attention due to its unique properties
\cite{Graphene_Review}.  In addition to the significant interest in
fundamental physics, which stems in part from the relativistic-like
behavior of charge carriers, this material is considered very
promising in many applications. Particularly, the possibility of
manipulating the band gap in a bilayer form \cite{Castro} offers an
additional control for nonlinear functionality.

While there has been a large number of reports (both experimental
and theoretical) on graphene and its derivatives
\cite{MLG_Hwang,Fuhrer2009,BLG_Hwang,Finland},  most of the studies
on electron transport properties have concentrated on monolayer
graphene (MLG).  Among those requiring further investigation in
bilayer graphene (BLG), intrinsic carrier-phonon scattering is one
of the most crucial as it determines the ultimate limit of any
electronic applications.  Due to the nonpolar nature of the
material, the deformation potential approximation is commonly used.
However, the deformation potential constant that quantifies the
strength of electron-phonon coupling, must be determined outside the
developed theory, in the form of an empirical fitting parameter. The
experimental results lie in a broad range (e.g., $D = 10-50$ eV)
\cite{MLG_Hwang,Finland} and further improvement of accuracy is
essential for reliable analysis of transport characteristics.  The
{\it ab initio} numerical methods provide an alternative approach in
addressing this problem as demonstrated successfully in MLG
\cite{Lazzeri,Borysenko}.  In this study, we conduct a first
principles calculation of electron-phonon interaction in BLG with AB
(also known as Bernal) stacking. Specifically, the density
functional perturbation theory (DFPT) is used to obtain the crystal
phonon spectra and electron-phonon coupling matrix elements. The
results illustrate the differences with MLG, indicating significant
discrepancies in transport properties.

Although it has been reported that the electron-phonon matrix
elements in MLG can be estimated with an excellent accuracy using
the GW approximation \cite{Lazzeri} for the phonon modes associated
with the observed Kohn anomalies \cite{Kohn}, a considerably longer
(compared to the DFPT) calculation time that the GW method requires,
as well as the fact that the accuracy of DFPT is sufficient in many
applications including carrier transport, justifies the use of DFPT
for calculating electron-phonon matrix elements in BLG. We also
demonstrate that the Kohn anomalies that lead to the largest
discrepancy between DFPT and GW results, do not play an important
role in the intrinsic electron scattering in BLG and so the accuracy
of the calculated matrix elements is not essential in this case.

\begin{center}
\begin{figure}[tbp]
\includegraphics[bb= 4 4 418 285,width= 3in]{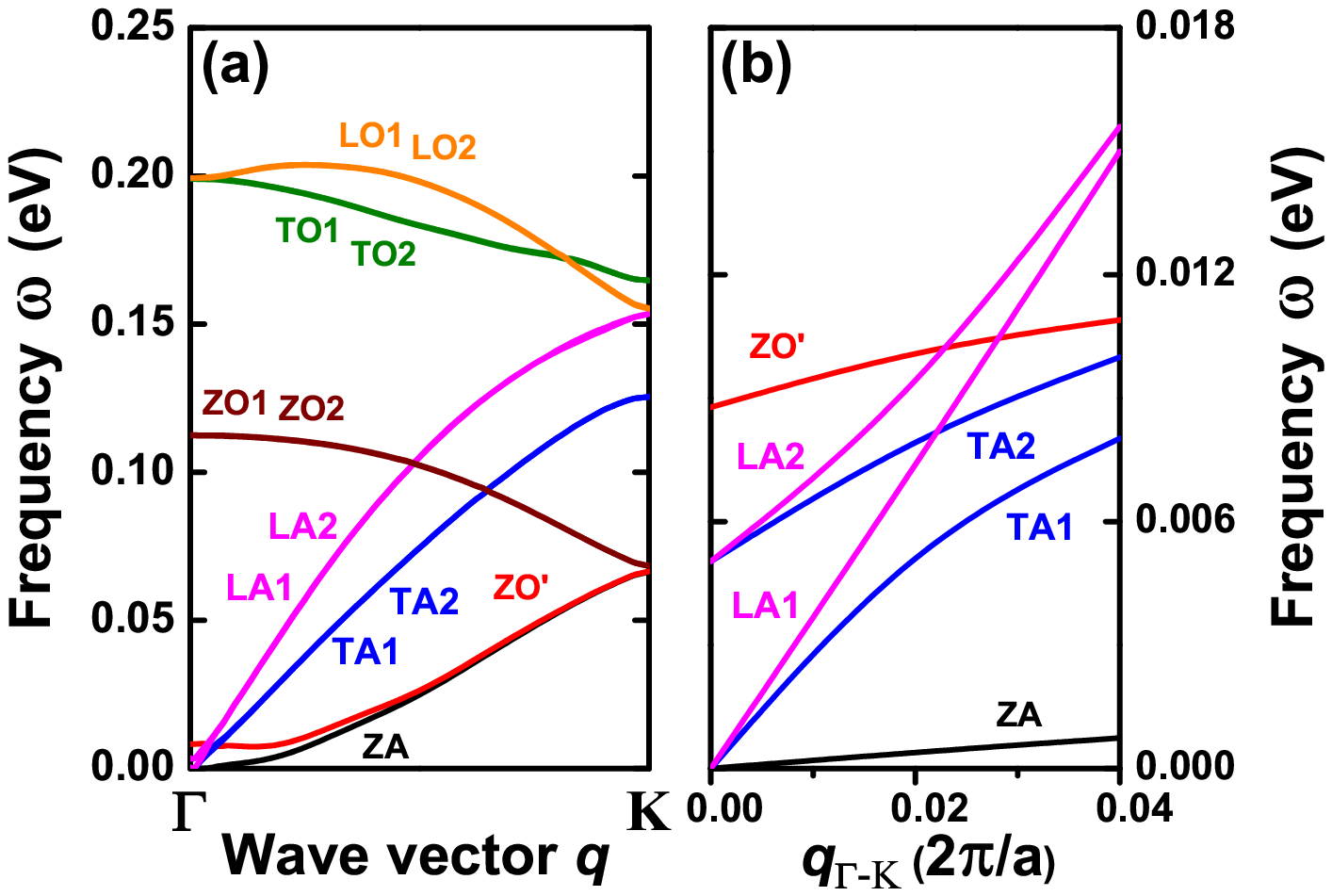}
\caption{(Color online) The phonon spectrum of BLG along the line
$\Gamma-K$, obtained by DFPT method (a). There are total of 12
phonon branches in BLG - twice as many as in MLG. But the weak van
der Waals interlayer coupling results in effective double degeneracy
of every branch, everywhere in FBZ except the vicinity of
$\Gamma$-point where a small splitting is exhibited in case of first
six branches (b). The largest splitting occurs between two
out-of-plane branches: ZA and ZO$^\prime$ ("breathing mode").}
\label{FIG_Phonon_Dispersion}
\end{figure}
\end{center}

Figure~\ref{FIG_Phonon_Dispersion} shows the calculated phonon
dispersion $\omega^{\nu}_\mathbf{q}$ along the $\Gamma$-$K$
direction. Since there are four carbon atoms in the unit cell, BLG
has 12 phonon branches.  However, only six lines are distinguishable
across the entire first Brillouin zone (FBZ) as each of them are
doubly degenerate due to the weak van der Waals coupling between the
two carbon layers.  The only exception, when the splitting has an
appreciable magnitude, is the case of ZA (acoustic out-of-plane) and
ZO$^\prime$ ("layer breathing") modes near the FBZ center. In
addition, two other modes deviate from the pure acoustic nature with
very small non-zero frequencies at the $\Gamma$ point as shown in
Fig.~\ref{FIG_Phonon_Dispersion}(b) (denoted as TA2 and LA2). Hence,
only three branches are truly acoustic (ZA, TA1, LA1), while we will
use a term "antisymmetric acoustic" to refer to the three
low-frequency counterparts with acoustic-like behavior (ZO$^\prime$,
TA2, LA2).  Their polarization eigenvectors are akin to those of the
acoustic modes in MLG, with the atomic displacements in two layers
having opposite phase (hence "antisymmetric").  A similar
terminology has been used  in Raman spectroscopy studies
\cite{Castro_Neto_Raman}.

\begin{center}
\begin{figure}[tbp]
\includegraphics[bb=80 84 539 692,width= 3in]{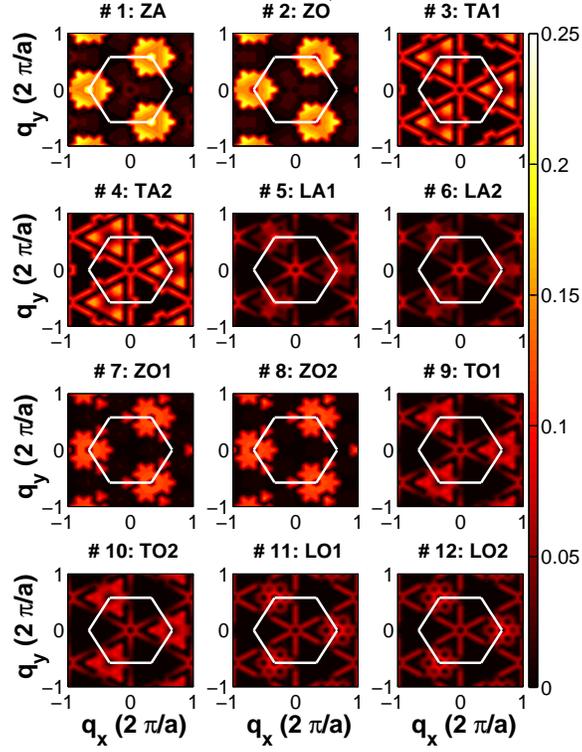}
\caption{(Color online) Intraband ($\pi^*_1$) matrix elements
$\left| g^{(\pi^*_1,\pi^*_1) \nu}_{\mathbf{k} + \mathbf{q},
\mathbf{k}} \right|$ (in units of eV) calculated by DFPT for
$\mathbf{k}$ at the conduction band minimum (i.e., the Dirac point)
as a function of phonon wavevector $\mathbf{q}$ and branch number
$\nu$. If compared to the matrix elements of MLG, the main
difference is the absence of Kohn anomalies - sharp peaks in
$\textbf{q} = \textbf{K}$ (TO1, TO2) and $\textbf{q} = 0$ (LO1,
LO2). } \label{FIG_g_DFPT}
\end{figure}
\end{center}

\begin{center}
\begin{figure}[tbp]
\includegraphics[bb= 3 228 585 524,width= 3in]{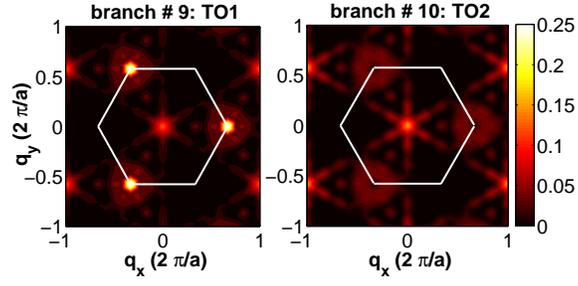}
\caption{Interband matrix elements ($\pi^*_1 \rightarrow
\pi^*_2$) (in units of eV) for phonon branches TO1 and TO2
($\textbf{k} = K$). The Kohn anomalies are clearly visible in a form
of sharp peaks. The observed picture agrees with the selection rules
for electron-phonon interaction in
BLG.~\cite{Malard}} \label{FIG_g_DFPT_CB12}
\end{figure}
\end{center}

The electron spectrum $E^i_{\mathbf{k}}$ of BLG has two close
conduction bands ($\pi^*_1$ and $\pi^*_2$); the distance between the
bottoms of $\pi^*_1$ and $\pi^*_2$ is found to be in a range $0.35 -
0.4$ eV \cite{Castro_Neto_Raman,PKim,Piscanec}. The accurate
estimate of the intrinsic electron scattering in BLG requires taking
into account all four possible scattering scenarios: intraband
($\pi^*_1 \rightarrow \pi^*_1$, $\pi^*_2 \rightarrow \pi^*_2$), and
interband ($\pi^*_1 \rightleftarrows \pi^*_2$). Correspondingly, we
calculate four sets of matrix elements $|g^{(i,j) \nu}_{\mathbf{k} +
\mathbf{q}, \mathbf{k}}|$ that represent the quantum mechanical
probability of electron transition from the state $(i,\mathbf{k})$
to $(j,\mathbf{k} + \mathbf{q})$, where $i$ and $j$ are the initial
and final energy band respectively. The matrix element is calculated
only for wave vectors $\textbf{q}$ in the irreducible Brillouin
zone. However it is easy to obtain the matrix elements in the entire
FBZ, by restoring the star of each $\textbf{q}$-vector from the
irreducible zone (which requires applying corresponding symmetry
operations to the vectors $\textbf{q}$ and $\textbf{k}$
simultaneously). The result of this calculation for intraband
scattering (when the electron is at the bottom of $\pi^*_1$, i.e.
$\textbf{k} = K$) is shown in Fig.~\ref{FIG_g_DFPT}. It is not
surprising that MLG-like phonon modes (ZA, TA1, LA1, ZO1, TO1, LO1)
and their BLG-counterparts (ZO$^\prime$, TA2, LA2, ZO2, TO2, LO2)
have very similar electron-phonon matrix elements throughout FBZ.
The difference is only exhibited in some cases, in points of high
symmetry ($\textbf{q} = \textbf{K}$ and $\textbf{q} = 0$). This can
be explained by selection rules that apply to phonon-mediated
electron transitions, and it results from different symmetries of
atomic displacements of these two sets of modes \cite{Malard}.  As
compared to MLG, the matrix elements have the same order of
magnitude. The essential difference between matrix elements of MLG
and the ones shown in Fig.~\ref{FIG_g_DFPT} is that none of the
in-plane optical phonon modes of BLG reveal Kohn anomalies that were
reported to be observed in a form of singularities at points
$\textbf{q} = \textbf{K}$ and $\textbf{q} = 0$ \cite{Borysenko}.
However, further calculations have shown that strong peaks similar
to those found in MLG occur in the corresponding matrix elements of
\emph{interband} scattering $\pi^*_1$ $\rightarrow$ $\pi^*_2$.
Figure~\ref{FIG_g_DFPT_CB12} shows the matrix element of interband
scattering $\pi^*_1 \rightarrow \pi^*_2$ due to the phonons of
branches 9 and 10 (TO1 and TO2). The symmetry analysis of these two
branches imposes a selection rule that allows only participation of
one of the two phonons with $\textbf{q} = \textbf{K}$ \cite{Malard}.
In agreement with this prediction, the sharp peaks at $\textbf{q} =
\textbf{K}$ are observed only in the case of TO1. At the same time,
the other Kohn anomaly (at $\textbf{q} = 0$) is present in matrix
elements of all four MLG-like optical branches: TO1, TO2, LO1, LO2
(see Fig.~\ref{FIG_g_DFPT_CB12}).

Based on the obtained matrix elements, the electron scattering rates
are calculated using Fermi's golden rule
\begin{equation} \left(\frac{1}{\tau}\right)_\mathbf{k}^{(i,j)
\nu} = \frac{2\pi}{\hbar} \sum_{\mathbf{q}} \left|g^{(i,j)
\nu}_{\mathbf{k} + \mathbf{q}, \mathbf{k}}\right|^2 \Delta^{(i,j)
\nu}_{\mathbf{k},\mathbf{q}},
 \label{Scatt_Rate}
\end{equation}
where $\Delta^{(i,j) \nu}_{\mathbf{k},\mathbf{q}} =
\left(N^{\nu}_\mathbf{q} + \frac{1}{2} \pm
\frac{1}{2}\right)\delta\left(E^j_{\mathbf{k} + \mathbf{q}} -
E^i_{\mathbf{k}} \pm \hbar\omega^{\nu}_\mathbf{q}\right)$,
$N^{\nu}_\mathbf{q}$ is the phonon population factor, and the sign
plus (minus) corresponds to the emission (absorption) of the phonon
$\omega^{\nu}_\mathbf{q}$. The temperature dependence of the
scattering rate stems only from the the numbers of phonons as we sum
over wave vectors $\textbf{q}$ in the FBZ. Preliminary estimates
confirm that the scattering rate, as a function of the initial
electron state $\textbf{k}$ in the vicinity of the Dirac point
$\textbf{K}$, exhibits isotropy. Therefore we only consider the
$K-\Gamma$ direction in $\textbf{k}$-space and can represent
$\tau^{-1}$ as a function of electron energy.

In the integration over FBZ the tight binding approximation  was
used to describe the electron energy spectrum $E_{\mathbf{k}}$ in
BLG \cite{Graphene_Review}. The second conduction band ($\pi^*_2$)
has nearly the same shape as the lowest conduction band ($\pi^*_1$)
near Dirac point $\textbf{K}$: it is parabolic in close vicinity of
$\textbf{k} = \textbf{K}$ and quasi-linear beyond that. The
quasi-linear approximation holds true for quite high electron
energies and therefore, this description appears sufficient in most
calculations. In the tight binding approximation, the distance
between the bottoms of $\pi^*_1$ and $\pi^*_2$ is $\gamma_1$, which
is the interlayer hopping integral in BLG (we use standard notations
known as the Slonczewski-Weiss-McClure parametrization model
\cite{Slonczewski,McClure}). A recent resonant Raman studies show
the result $\gamma_1 \simeq$ 0.4 eV \cite{PKim,Piscanec}, which is
the value used in our calculation.

\begin{center}
\begin{figure}[tbp]
\includegraphics[bb=2 2 208 137,width=3in]{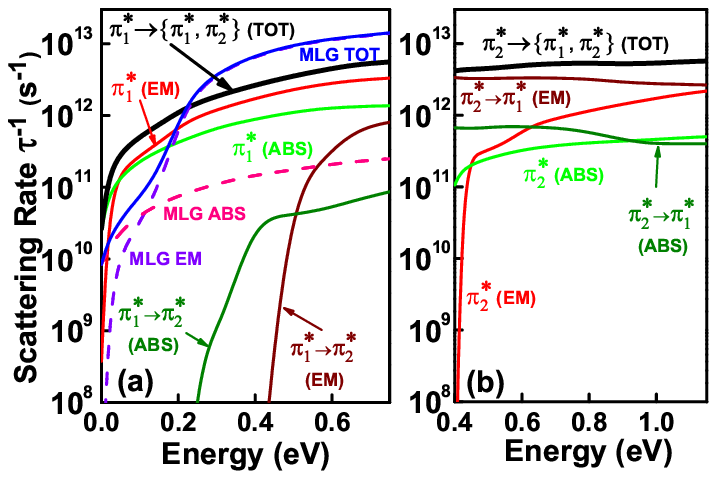}
\caption{(Color online) Scattering rates at $T = 300 $ K as a
function of electron energy $E_\mathbf{k}$, in BLG. The results are
rather different depending on whether the electron is initially in
the conduction band $\pi^*_1$ (a), or $\pi^*_2$ (b). In the former
case, the dominance of the intraband ($\pi^*_1$) scattering is
clear. The scenario when the electron is initially in $\pi^*_2$
becomes more relevant when the gate bias is applied.}
\label{FIG_R_DFPT}
\end{figure}
\end{center}

Figure~\ref{FIG_R_DFPT} shows the total scattering rates associated
with all four possible scenarios involving two conduction bands.
Since the recent Raman experiments in BLG have shown the Kohn
anomalies located in the q-points of high symmetry, one might expect
their significance in the overall picture of electron-phonon
scattering in BLG. The separate analysis of different phonon
branches in BLG, however, shows a qualitatively different picture.
In case of intraband scattering in the lowest conduction band
($\pi^*_1$) the scattering rate due to absorption or emission of
MLG-like optical phonons (ZO1, ZO2, TO1, TO2, LO1, LO2) is
negligibly small in a wide range of energies ($E < 0.7$ eV). An
explanation for this result comes directly from comparison of
corresponding matrix elements of MLG \cite{Borysenko} and those of
BLG for $\pi^*_1 \rightarrow \pi^*_1$ scattering: in the latter case
the sharp peaks are missing. The Kohn anomalies similar to those in
matrix elements of MLG, were observed in BLG - in the interband
scattering $\pi^*_1 \rightarrow \pi^*_2$ (see
Fig.~\ref{FIG_g_DFPT_CB12}). However, the corresponding scattering
rates have proven to be negligibly low. This result can be
understood if one compares electron dispersions in MLG ($\pi^*$) and
BLG ($\pi^*_1$ and $\pi^*_2$). A simple analysis demonstrates that
electron transitions $\pi^*_1 \rightarrow \pi^*_2$ involving phonons
$\textbf{q} = 0$ and $\textbf{q} = \textbf{K}$ (where the anomalies
occur) are prohibited by the energy and momentum conservation laws.
In other words, the contour of integration from
Eq.~(\ref{Scatt_Rate}) never crosses (or comes close to) the peaks
at the points $\Gamma$ and $K$ observed in the matrix element of
BLG. The estimated minimal distance between this curve (which can be
approximated by a circle) and the point $\textbf{q} = \textbf{K}$,
is 0.034 $\left(2 \pi/a\right)$ for emission and 0.014 $\left(2
\pi/a\right)$ for absorption. For comparison, the analogous distance
in case of intervalley scattering due to TO phonons in MLG, is much
smaller, $\sim 10^{-7}\left(2 \pi/a\right)$, which ensures the
dominant role of the corresponding modes in electron scattering. The
same consideration shows that the TO1 phonon with $\textbf{q} = 0$
cannot influence the interband scattering in BLG. As a result, the
total scattering rate in BLG, when the electron is initially in
$\pi^*_1$ (i.e., $\pi^*_1$$\rightarrow$\{$\pi^*_1$, $\pi^*_2$\}) is
somewhat smaller at higher energies ($E_k \gtrsim 200$ meV), if
compared to the total rate in MLG (see Fig.~\ref{FIG_R_DFPT}).

On the other hand, the low-energy scattering rate in BLG is
substantially higher, which can be explained by a larger density of
states in the vicinity of the Dirac point.  More specifically, the
density of state tends to zero linearly for the massless Dirac
fermions of MLG while it is a constant in BLG as
$\textbf{k}\rightarrow K$. This non-zero density of states at the
bottom of $\pi^*_1$ and $\pi^*_2$ leads to a peculiar effect. In
MLG, the intravalley scattering on acoustic phonons can be treated
as quasielastic as it was shown previously \cite{Vasko,Hwang_08};
only the modes with $\textbf{q}\rightarrow 0$ can participate. As a
result, the scattering rate is a linear function of energy that
vanishes as $E_k \rightarrow 0$. In BLG, however, the intravalley
scattering with absorption of acoustic phonon shows a different
behavior. Aside from $\textbf{q} = 0$, the energy and momentum
conservation in the electron scattering allows an additional
solution - a circle with radius $\left|\textbf{q}\right| = q_c = v_s
\gamma_1/\hbar v_F^2$. In the case of antisymmetric acoustic modes,
an equivalent solution exists - a circle of radius $q_c =
\sqrt{\gamma_1 \omega_{\Gamma}/\hbar v_F^2}$, where
$\omega_{\Gamma}$ is the non-zero frequency of an antisymmetric
acoustic mode in $\Gamma$-point. This leads to a non-zero absorption
rate even at zero electron energies. It is important to emphasize
that even though the deformation potential approximation can be used
to describe electron-phonon coupling in BLG \cite{Finland}, the
intravalley scattering cannot be treated as quasi-elastic due to the
additional solution with $\textbf{q} \neq 0$. The estimated value of
the deformation potential constant in BLG is $D \approx 15$ eV, if
we chose the TA2 branch as dominant \cite{comment_DPA}. This is much
closer to the value obtained for graphite ($ D \approx 16$ eV
\cite{Sugihara}) than that for MLG ($D \approx 4.5-5$ eV
\cite{Borysenko}).  The scattering rates in BLG and graphite
\cite{Dresselhaus} turn out to be quite similar as well. The same
dominant intrinsic scattering mechanism in graphite and BLG is
intravalley transitions due to absorption and emission of acoustic
phonons, while the contribution of the optical phonons is of
secondary importance. In particular, $D_{op} \approx
2.8\times10^8$~eV/cm for the optical phonons in BLG, which is almost
one order of magnitude smaller than in MLG
\cite{comment_DPA,Borysenko}. With much enhanced acoustic phonon
scattering at low energies, the low field mobility of bilayer
graphene is expected to be significantly smaller than that of
monolayer graphene.

In summary, we have studied the electron-phonon interaction in BLG
with AB stacking and found that the scattering rates exhibit
qualitatively different behavior, if compared to MLG. As a result of
a different electron spectrum (and thus, the density of states) in
conduction bands of BLG, the intraband ($\pi^*_1$) scattering rates
due to absorption of acoustic and antisymmetric acoustic phonons at
low electron energies have quite large values ($\sim10^{9}-10^{10}$
s$^{-1}$). The calculations confirmed that this scattering mechanism
remains dominant in BLG, at all relevant electron energies in
$\pi^*_1$ --- the property that is not observed in MLG. In
$\pi^*_2$, the scattering rate is approximately a constant, $\approx
5 \times 10^{12}$ s$^{-1}$. The overall picture of electron-phonon
coupling in BLG has more in common with that of the bulk graphite
than it does with MLG. The estimated values of the deformation
potential constants are $D \approx 15$ eV and $D_{op} \approx
2.8\times10^8$~eV/cm, for acoustic and optical phonons respectively.

This work was supported, in part, by the DARPA/HRL CERA, ARL, and
SRC/FCRP FENA programs. MBN wishes to acknowledge partial support
from the Office of Basic Energy Sciences, US DOE at Oak Ridge
National Lab under contract DE-AC05-00OR22725 with UT-Battelle, LLC.
JMZ acknowledges support from NSF under the IR/D program.

\newpage

\newpage

\newpage

\newpage


\begin{thebibliography}{99}
\bibitem{comment_DPA} The best fitting results for the total
scattering rate are achieved when the antisymmetric acoustic branch
TA2, and optcal branch TO2 are chosen, with the effective
deformation potential constants $D \approx 15$~eV and $D_{op}
\approx 2.8\times10^8$~eV/cm, respectively. There are six
acoustic-like (ZA, ZO$^{\prime}$, TA1, TA2, LA1, LA2) and four
optical (TO1, TO2, LO1, LO2) vibrational modes in BLG that
contribute almost equally in a wide range of energies. Thus, the
branches TA2 and TO2 were chosen simply to ensure the best fit, and
not because of their exceptional role.
\end{thebibliography}
\end{document}